\documentclass[12pt]{article}
\def\beq{\begin{equation}}
\def\eeq{\end{equation}}
\def\ap#1#2#3 {Ann. Phys. (NY) {\bf#1} (19#2) #3}
\def\err#1#2#3 {{\it Erratum} {\bf#1} (19#2) #3}
\def\ib#1#2#3 {{\it ibid.} {\bf#1} (19#2) #3}
\def\ijmp#1#2#3 {Int. J. Mod. Phys. {\bf#1} (19#2) #3}
\def\jetp#1#2#3 {JETP Lett. {\bf#1} (19#2) #3}
\def\mpl#1#2#3 {Mod. Phys. Lett. {\bf#1} (19#2) #3}
\def\np#1#2#3 {Nucl. Phys. {\bf#1} (19#2) #3}
\def\pl#1#2#3 {Phys. Lett. {\bf#1} (19#2) #3}
\def\prep#1#2#3 {Phys. Rep. {\bf#1} (19#2) #3}
\def\prev#1#2#3 {Phys. Rev. {\bf#1} (19#2) #3}
\def\prl#1#2#3 {Phys. Rev. Lett. {\bf#1} (19#2) #3}
\def\sjnp#1#2#3 {Sov. J. Nucl. Phys. {\bf#1} (19#2) #3}
\def\spj#1#2#3 {Sov. Phys. JETP {\bf#1} (19#2) #3}
\def\spu#1#2#3 {Sov. Phys. Usp. {\bf#1} (19#2) #3}
\def\zp#1#2#3 {Zeit. Phys. {\bf#1} (19#2) #3}

\begin{document}
\begin{titlepage}
\begin{center}
{\Large \bf Theoretical Physics Institute \\
University of Minnesota \\}  \end{center}
\vspace{0.2in}
\begin{flushright}
TPI-MINN-00/52-T \\
UMN-TH-1928-00 \\
November 2000 \\
\end{flushright}
\vspace{0.3in}
\begin{center}
{\Large \bf  Decays of $b$ hadrons and a possible new four-quark
interaction
\\}
\vspace{0.2in}
{\bf M.B. Voloshin  \\ }
Theoretical Physics Institute, University of Minnesota, Minneapolis,
MN
55455 \\ and \\
Institute of Theoretical and Experimental Physics, Moscow, 117259
\\[0.2in]
\end{center}

\begin{abstract}

A possibility is considered of explaining the low experimental value of
the ratio of the lifetimes $\tau(\Lambda_b)/\tau(B_d)$ by a new
`centiweak' four-quark interaction, i.e with a strength on the order of
$10^{-2} \, G_F$.  It is noted that the considered interaction can also
improve agreement with the data on low semileptonic branching ratio
$B_{sl} (B)$ in $B$ meson decays with a simultaneous slight decrease in
the prediction for the average charm yield in those decays. The proposed
new interaction modifies within the present experimental limits the
predictions for differences of lifetimes among $B$ mesons, and can thus
be probed by more precise data on these differences. A sample model is
briefly discussed, where the new interaction arises through a weak SU(2)
singlet scalar field with quantum numbers of a diquark.
\end{abstract}


\end{titlepage}

\section{Introduction}
The central value of the experimental data on the ratio of the lifetimes
\cite{pdg} $\tau(\Lambda_b)/\tau(B_d) = 0.79 \pm 0.05$ persistently
defies an explanation within the present theoretical understanding of
differences of inclusive decay rates of heavy hadrons in the Standard
Model, which is highly unlikely to produce a theoretical prediction for
this ratio outside the range $\tau(\Lambda_b)/\tau(B_d) > 0.9$ (for a
review see e.g. Ref.\cite{mv}, and also the most recent review \cite{nu}
). If this conundrum is not resolved by improved experimental data, we
might be compelled to look for an explanation beyond the Standard Model.
In fact, the decays of $b$ hadrons are a likely place where effects of
new physics may show up. Indeed, the dominant weak interaction in these
decays is suppressed by $|V_{cb}| \approx 0.04$, thus new interactions
with a `centiweak' strength of about $10^{-2} \, G_F$ may lead to
sizeable relative effects in the $b$ decays. This paper hypothesizes on
an impact on these decays of one such `centiweak' four-quark interaction
with the chiral structure
\beq
({\overline c}_L \, \Gamma \, b_R) ({\overline d}_L \Gamma^\prime u_R)
~,
\label{cwi}
\eeq
where $\Gamma$ and $\Gamma^\prime$ generically stand for spin and color
matrices. The presence of a right-chiral $u$ quark prevents an
interference between the new hypothetical term and the weak interaction
in the `parton' decay of the $b$ quark, so that the new interaction
contributes only quadratically to the overall total non-leptonic decay
rate. On the other hand, the chirality of light quarks is broken in
hadrons, so that the differences of inclusive decay rates of individual
hadrons, arising from the effects of a spectator $u$ quark receive a
linear contribution from the new interaction through its interference in
the spectator effects with the standard weak interaction. Thus the
impact of the new interaction on the differences of the inclusive decay
rates can be comparable with their values in the standard calculation,
while the effect in the overall non-leptonic decay rate would amount to
only about  15 \%.
It can be noted that an increase of the non-leptonic `parton' decay rate
of such magnitude is not unwelcome in view of the long-standing problem
(see e.g. in Refs. \cite{bbsv,bsu,nu}) of a somewhat low semileptonic
branching ratio $B_{sl}(B)$.

In what follows it will be shown, that with the strength of the new
interaction corresponding to such moderate increase in the `parton'
non-leptonic decay rate, one can achieve an enhancement of the
$\Lambda_b$ decay rate from an additional contribution to the `weak
scattering' $u \, b \to c \, d$. The specific estimate of the
enhancement is, as usually, complicated by the uncertainty in hadronic
matrix elements of four-quark operators. For reasonable values of those
matrix elements a `sample' estimate of the enhancement amounts to
approximately 5\%. Although the magnitude of the enhancement does not
look excessively large by itself, one should
take into consideration that it essentially equals to the calculated
effect within the Standard Model. Simultaneously with this enhancement
of the $\Lambda_b$ decay, the new interaction additionally suppresses
(within the current experimental errors) the inclusive decay rate of the
$B^-$ meson through an additional contribution to the negative `Pauli
interference'. Thus a larger, than conventionally predicted, ratio of
the lifetimes $\tau(B^-)/\tau(B^0)$, may be viewed as a test of the
considered mechanism with a new interaction. The connection between the
enhancement of the $\Lambda_b$ decay and the suppression of that of the
$B^-$ could be avoided if the suggested interaction included a
right-chiral $d_R$ quark rather than $u_R$. The reason to choose the
structure as shown in eq.(\ref{cwi}) is that it can be realized by
exchange of a colored scalar field $\phi$:
$b_R \, u_R \to \phi \to c_L \, d_L$, which is a singlet under the
electroweak SU(2) symmetry. Thus as `unnatural' as it would be, it is
perfectly consistent to consider the situation where only the quark
flavors indicated in eq.(\ref{cwi}) are involved in a new interaction,
mediated by $\phi$. In this case, to the best of the knowledge of the
author, an interaction of this type with a centiweak strength does not
contradict the known phenomenology of the flavor dynamics.
On the contrary, an interaction involving a right-chiral $d$ quark
instead of $u_R$ would arise from an exchange of a scalar, which is a
component of an SU(2) doublet, and one would have to deal with
potentially unwanted consequences of the interactions induced by the
other component of the doublet.

In what follows only the flavor structure as shown in eq.(\ref{cwi}) is
considered, containing all four possible spin and color combinations. In
Sect.2 the expressions are derived for the effects of the suggested
interaction in the `parton' decay rate of the $b$ quark, and in the
spectator effects in $\Lambda_b$ and in $B$ mesons. The numerical
estimates of these effects are presented in Sect.3, followed by Sect.4,
containing the discussion and summary.

\section{Effects of the new interaction in decays of $b$ hadrons}
The Lagrangian for the hypothetical new interaction with the chiral
structure shown in eq.(\ref{cwi}) can be written in the following
general form
\begin{eqnarray}
&&L_{cw}={G_F \, V_{cb} \over \sqrt{2}} \, \left [ \right. 2 \, h_1 \,
(\overline c \, (1+ \gamma_5) \, b )  ( \overline d \, ( 1+ \gamma_5) \,
u  ) + {1 \over 2} \, h_2 \,  (\overline c \, \sigma_{\mu \nu} \,(1+
\gamma_5) \, b  )  ( \overline d \, \sigma^{\mu \nu} \,( 1+ \gamma_5) \,
u  )
\nonumber \\
&&+2 \, h_3 \,  (\overline c \, (1+ \gamma_5) \, u  )  ( \overline d \,
( 1+ \gamma_5) \, b ) + {1 \over 2} \, h_4 \,  (\overline c \,
\sigma_{\mu \nu} \,(1+ \gamma_5) \, u  )  ( \overline d \, \sigma^{\mu
\nu} \,( 1+ \gamma_5) \, b  ) \left. \right ] + h.c.~,
\label{lcw}
\end{eqnarray}
where $\sigma_{\mu \nu}=i \, (\gamma_\mu \gamma_\nu - \gamma_\nu
\gamma_\mu)/2$, and the overall factor $G_F \, V_{cb} / \sqrt{2}$ is
chosen in such a way that the dimensionless constants $h_1, \ldots ,
h_4$ describe the strength of the interaction relative to the dominant
four-quark interaction in nonleptonic $b$ decays:
\begin{eqnarray}
&&L_{w}={G_F \, V_{cb} \over \sqrt{2}} \, \left [ \right. {C_+ + C_-
\over 2} \,  (\overline c \, \gamma_\mu \, (1- \gamma_5) \, b )  (
\overline d \, \gamma_\mu ( 1- \gamma_5) \, u  ) +
\nonumber \\
&&{C_+ - C_- \over 2} \,  (\overline c \, \gamma_\mu \, (1- \gamma_5) \,
u  )  ( \overline d \, \gamma_\mu ( 1- \gamma_5) \, b )  \left. \right ]
+ h.c.~,
\label{lw}
\end{eqnarray}
with $C_\pm$ being the well-known short-distance renormalization
coefficients:
\beq
C_-=C_+^{-2}=\left [ {\alpha_s(m_b) \over \alpha_s(m_W)}
\right ]^{4/\beta_0}~,
\label{cpm}
\eeq
and $\beta_0$ is the coefficient in the QCD beta function. The value of
$\beta_0$
relevant to $b$ decays is $\beta_0=23/3$. The terms in both the standard
weak interaction Lagrangian (\ref{lw}) and the hypothetical one in
eq.(\ref{lcw}) are assumed to be normalized at $\mu=m_b$, so that the
constants $h_A$ as well as $C_\pm$ stand for their values at this
normalization point: $h_A=h_A(m_b)$, $C_\pm=C_\pm(m_b)$, and the
appropriate normalization of the four-quark operators is also implied.

\subsection{Effect in the total nonleptonic decay rate}
As is already discussed, due to extremely small mass of the $u$ quark,
there is essentially no interference between the `centiweak' and the
standard weak decay amplitudes in the total `parton' decay rate of the
$b$ quark. The effect of the new interaction is thus quadratic in the
coefficients $h_A$, and is given by
\beq
{\delta \Gamma_{nl}(b) \over \Gamma_{nl}^{(0)}(b)}= |h_1|^2 + 3 \,
|h_2|^2 + |h_3|^2 +3 \, |h_4|^2+ {2 \over 3} \, {\rm Re} \left ( h_1 \,
h_3 ^*+3 h_2 \, h_4^* \right )~,
\label{dgtot}
\eeq
where $ \Gamma_{nl}^{(0)}(b)$ is the standard `parton'  nonleptonic
decay rate for the process $b \to c \, d \, {\overline u}$:
\beq
 \Gamma_{nl}^{(0)}(b)= {G_F^2 \, |V_{cb}|^2 m_b^5 \over 64 \, \pi^3}
\left [ (1-{m_c^4 \over m_b^4})(1-8\, {m_c^2 \over m_b^2}+ {m_c^4 \over
m_b^4})-24 \, {m_c^4 \over m_b^4} \, \ln {m_c \over m_b} \right ]~,
\label{gstand}
\eeq
with the QCD radiative corrections being omitted in both
eq.(\ref{dgtot}) and eq.(\ref{gstand}).

\subsection{Interference with standard weak interaction in decays of
$\Lambda_b$}

The current theoretical approach to the effects of spectator quarks in
inclusive decay rates of heavy hadrons is based on the operator product
expansion in powers of $m_Q^{-1}$ of the `effective Lagrangian'
corresponding to the absorptive part of the correlator:
\beq
L_{eff}=2 \,{\rm Im} \, \left [ i \int d^4x \, e^{iqx} \, T \left \{
L_{int}(x),
L_{int}(0) \right \} \right ]~,
\label{leff}
\eeq
where $L_{int}$ is the part of the (weak) interaction Lagrangian
responsible for the type of the decay under consideration. In terms of
$L_{eff}$ the inclusive decay rate of a heavy hadron $H_Q$ is given
by\footnote{We use here
the non-relativistic normalization
for the {\it heavy} quark states: $\langle
Q | Q^\dagger Q | Q \rangle =1$.}
\begin{equation}
\Gamma_H=\langle H_Q | \, L_{eff} \, | H_Q \rangle~.
\label{lgam}
\end{equation}

The leading term in the OPE describes the `parton' decay rate of a heavy
quark, which is the same for all hadrons carrying the given heavy quark
flavor, while the differences between decay rates of individual hadrons
are described by the terms relatively suppressed by $m_Q^{-2}$:
$L_{eff}^{(2)}$, and $m_Q^{-3}$: $L_{eff}^{(3)}$. In the standard
analysis (see e.g. in Ref.\cite{mv}) the contribution of
$L_{eff}^{(2)}$ to the difference of lifetimes of $\Lambda_b$ and $B^0$
amounts to only about 1\%, while the $L_{eff}^{(3)}$ term enhances the
decay rate of $\Lambda_b$ relative to $B$ by at most 6-7\%.

The hypothetical new interaction in eq.(\ref{lcw}) gives rise to a cross
term in the correlator (\ref{leff}) between $L_{w}$ and $L_{cw}$, giving
an additional contribution $\delta L_{eff}^{(3)}$ to $L_{eff}^{(3)}$.
For simplicity only the part of $\delta L_{eff}^{(3)}$ relevant to the
shift of the decay rates of the hyperons in the triplet ($\Lambda_b$,
$\Xi_b^0$, $\Xi_b^-$) is written here. In separating out this part one
takes into account that in these baryons there is no correlation of the
heavy quark spin with the spin degrees of freedom of the light quarks.
Technically this reduces to the following replacement for the bilinear
in the heavy quark operator matrix:
\beq
{\overline b}_{i \, \alpha} b^{j \, \beta} \to \left ( {1+ \gamma^\mu
u_\mu \over 4} \right)^\beta_\alpha \, \left ( {\overline b}_{i} b^{j}
\right )~,
\label{substb}
\eeq
where $\alpha, \beta$ are Dirac spinor indices, $i, \, j$ are the color
indices, and $u_\mu$ is the 4-velocity of the baryon.
Furthermore, the average value of the parity-violating operators over
the hadrons are vanishing, thus the only relevant structures reduce to
those containing the operators $({\overline b} b)({\overline u} u)$ and
$({\overline b}_{i} b^{j} ) ({\overline u}_{j} u^{i} )$. Also the
`hybrid' renormalization factors are not included here, following the
convention
that the operators, as well as the constants $h_a$ in the new
interaction (\ref{lcw}) are normalized at $\mu=m_b$. After these
remarks, the relevant part of the effective Lagrangian reads as
\begin{eqnarray}
&&\delta \Gamma (H_b) = \langle H_b |\delta L_{eff}^{(3)} | H_b \rangle
= {G_F^2 m_b^2
|V_{cb}|^2 \over 4 \pi} \times \nonumber \\
&&\left \{ \right .  {\rm Re}\left[ C_+ (h_1+ h_3-3 h_2 - 3 h_4) + C_-
(h_1 - h_3 - 3 h_2 + 3 h_4) \right ] \langle H_b | ({\overline b}
b)({\overline u} u) | H_b \rangle +   \nonumber \\
&&  {\rm Re}\left[ C_+ (h_1+ h_3-3 h_2 - 3 h_4) - C_- (h_1 - h_3 - 3 h_2
+ 3 h_4) \right ] \langle H_b | ({\overline b}_{i} b^{j} ) ({\overline
u}_{j} u^{i} ) | H_b \rangle  \left. \right \}~.
\label{iswwh}
\end{eqnarray}
In descriptive terms this expression corresponds to the additional
contribution from the interference of the new interaction with the
standard one to the weak scattering $b \, u \to c \, d$ in decays of the
$b$ hyperons.

\subsection{Interference in $B^-$}

The term in $L_{eff}^{(3)}$ with the four-quark operators of the type
$({\overline b} \, \Gamma \, u)({\overline u} \, \Gamma^\prime \, b)$
corresponding to the weak scattering in decays of the hyperons, when
applied to mesons, describes the Pauli interference of the spectator
${\overline u}$ antiquark in $B^-$ meson with that produced in the decay
$b \to c \, d \, {\overline u}$. The expression for the effect of the
new interaction in this term is written here in the limit of
factorization, which in practice amounts to the following substitution:
\beq
\langle B^- |({\overline b} \, \Gamma \, u)({\overline u} \,
\Gamma^\prime \, b)|   B^- \rangle \to -{f_B^2 \, m_b \over 288} \, {\rm
Tr}\left[ \Gamma \, \gamma_5 \, (1+ \gamma \cdot u) \right]\, {\rm
Tr}\left[ \Gamma^\prime \, (1+ \gamma \cdot u)  \, \gamma_5 \right]
\label{fsub}
\eeq
with arbitrary spin and color matrices $\Gamma$ and $\Gamma^\prime$, and
with the traces running over the spinor and color indices. The quantity
$f_B$ in this expression is the $B$ meson annihilation constant.

After these preliminary remarks, the expression for the shift of the
total decay rate of $B^-$ due to the interference of the new and the
standard interaction is readily calculable, and is given by
\beq
\delta \Gamma(B^-)=-{G_F^2 m_b^3 f_B^2 |V_{cb}|^2 \over 4 \pi}\, {\rm
Re} \left [ C_+ \, ({1 \over 3} \, h_1 + h_3 - h_2 - 3 \, h_4) + C_- \,
({1 \over 3} \, h_1 - h_3 - h_2 + 3 \, h_4) \right ]
\label{dgbm}
\eeq

\subsection{Contribution of the new interaction to $B_d^0 \to c \,
{\overline u}$}
With the standard weak interaction the shift of the total decay rate of
the $B_d^0$ meson due to the annihilation process $b \, {\overline d}
\to c \, {\overline u}$ is suppressed by the factor $m_c^2/m_b^2$ for
chirality reasons. The hypothetical new interaction, involving $c$ and
$u$ of opposite chirality does not carry such suppression.  The
expression for the (necessarily positive) shift of the total decay rate
of $B_d^0$, quadratic in the new interaction, reads as
\beq
\delta \Gamma (B_d \to c \, \bar u)=|V_{cb}|^2 \,{G_F^2 \, m_b^3 \,
f_B^2 \over 8 \pi} \left |{1 \over 3} \, h_1 + h_2 -2 \, h_3 \right
|^2~.
\label{dgbd}
\eeq

\section{Estimates of the effects}
Clearly, the dependence of the discussed effects in $b$ decays on four
parameters $h_A$ leaves a considerable freedom for uncoupling the
magnitude of these effects in four specific processes. For a more
constrained discussion, it is assumed here that the new interaction
arises from an exchange of a scalar boson $\phi$ in the diquark channel:
$b_R \, u_R \to \phi \to c_L \, d_L$, which fixes the spinor
structure of the interaction, normalized at $\mu = m_\phi$. If $\phi$ is
a color (anti)triplet the relation between the constants at that
normalization point is $h_1(m_\phi)=h_3=3\, h_2= 3 h_4(m_\phi)$, while
if the $\phi$ is a color sextet, the corresponding relation is
$h_1(m_\phi)=h_3(m_\phi)=-h_2(m_\phi)=-h_4(m_\phi)$. In either case one
has $h_1=h_3$ and $h_2=h_4$ and this relation is preserved through the
QCD renormalization of the `centiweak' Lagrangian down to $\mu=m_b$. For
this reason in what follows the notation $h_S$ is used for the common
value of the coefficients $h_1$ and $h_3$ in front of the scalar
operators and $h_T$ stands for the common value of the coefficients
$h_2$ and $h_4$ for the tensor structures. We will not speculate here on
apriori relative values of $h_S$ and $h_T$, and
will leave these two constants (at $\mu=m_b$) as free parameters in the
present phenomenological analysis.

With this reduction in the number of parameters the expressions of the
previous section considerably simplify. In particular, the additional
shift of the total decay rate of the $B^-$ meson, given by
eq.(\ref{dgbm}), takes the form\footnote{In the numerical estimates here
we set $C_=1.35$, $C_+ = C_-^{-1/2} \approx 0.86$.}
\begin{eqnarray}
\delta \Gamma(B^-) &=&-{G_F^2 m_b^3 f_B^2 |V_{cb}|^2 \over 6 \pi}\, (2\,
C_+-C_-) {\rm Re} [h_S-3 \, h_T] \nonumber \\
&\approx& -0.03 \, {\rm Re}[h_S-3 \, h_T]
\, \left ( {f_B \over 200 \, MeV } \right )^2 \,
ps^{-1}~,
\label{dgbr}
\end{eqnarray}
and it should be emphasized again that this estimate assumes perfect
factorization of the four-quark matrix elements over the $B$ meson.

The formula in eq.(\ref{iswwh}) for the shift of the decay rate of
$\Lambda_b$ (and the same shift for $\Xi_b^0$) reduces to
\begin{eqnarray}
\delta \Gamma(\Lambda_b) &=& {G_F^2 m_b^2 |V_{cb}|^2 \over 2 \pi} \, C_+
{\rm Re} [h_S-3 \, h_T] \langle \Lambda_b | ({\overline b} b)({\overline
u} u)+ ({\overline b}_{i} b^{j} ) ({\overline u}_{j} u^{i} )|\Lambda_b
\rangle \nonumber \\
&\approx & 0.05 \, {\rm Re} [h_S-3 \, h_T] \left ( {\langle \Lambda_b |
({\overline b} b)({\overline u} u)+ ({\overline b}_{i} b^{j} )
({\overline u}_{j} u^{i} )|\Lambda_b \rangle \over 0.04 \, GeV^3 }
\right ) \, ps^{-1}~.
\label{dgaml}
\end{eqnarray}
The matrix element of the four quark operator involved in the latter
expression is a source of a major uncertainty. In a naive quark picture
the color antisymmetry of the quark wave function would lead to a
cancellation between the two terms differing by the `twist' of the quark
colors. However an analysis \cite{mv2,mv} of similar matrix elements
with a product of vector currents (rather than of the scalar densities):
$({\overline Q} \gamma_\mu Q)({\overline q} \gamma^\mu q)$ and
$({\overline Q}_{i}\gamma_\mu Q^{j} ) ({\overline q}_{j} \gamma^\mu
q^{i} )$, whose matrix elements are related to differences of lifetimes
within the triplet of charmed baryons ($\Lambda_c, \, \Xi_c^+, \,
\Xi_c^0$), reveals that the color antisymmetry relation badly fails, and
the value of $0.04 \, GeV^3$, used in eq.(\ref{dgaml}), is well
representative at least in the case of vector currents. In the case of
the operators with products of scalar densities as in eq.(\ref{dgaml})
we in fact have no good guidance\footnote{Except for the obvious remark
that in the heavy quark term $({\overline Q} \gamma^0 Q)$ is equivalent
to $({\overline Q} Q)$ in the leading in $m_Q$ approximation.}, and the
particular value, used in eq.(\ref{dgaml}) should be considered as an
`educated guess'.

The expression (\ref{dgtot}) for the additional nonleptonic `parton'
decay rate of a $b$ quark in terms of $h_S$ and $h_T$ reads as
\beq
{\delta \Gamma_{nl}(b) \over \Gamma_{nl}^{(0)}(b)}={8 \over 3} \left (
|h_S|^2 + 3 |h_T|^2 \right )~.
\label{dgtotm}
\eeq
Clearly the ratio $\delta \Gamma(\Lambda_b)/\delta \Gamma_{nl}(b)$ is
maximized, if one assumes that both $h_S$ and $h_T$ are real, $h_S$ is
positive, and $h_T=-h_S$\footnote{The latter relation is the one
corresponding to an interaction induced by an exchange of a color sextet
scalar $\phi$, if one neglects the QCD running effects.}. Taking as
sample values $h_S=-h_T=0.15$, and leaving aside the uncertainties in
the matrix elements involved in equations (\ref{dgbr}) and
(\ref{dgaml}), one estimates the discussed shifts of the decay rates
due to the new hypothetical interaction as $\delta \Gamma (\Lambda_b)
\approx 0.03 \, ps^{-1}$, $\delta \Gamma (B^-) \approx -0.02 ps^{-1}$,
and $\delta \Gamma_{nl}(b) \approx 0.24 \, \Gamma_{nl}^{(0)}(b)$. In
relative terms, these estimates correspond to an additional enhancement
of the $\Lambda_b$ decay rate by about 5\% of the decay rate of the
$B_d^0$ meson, a reduction of the $B^-$ decay rate by about 3 \%, and an
increase in the overall non-leptonic decay rate of the $b$ hadrons by
(12 -- 18)\%. The latter number is lower than the relative value of the
additional decay rate with respect to  $\Gamma_{nl}^{(0)}(b)$, due to
the contribution to the total nonleptonic rate of the decay $b \to c \,
{\overline c} \, s$.  The range of values corresponds to uncertainty in
the effect of yet uncalculated LLO and NLO QCD corrections to the $b$
decay rate through the discussed hypothetical interaction.

The additional contribution to the total decay rate of the $B_d^0$
meson, described by eq.(\ref{dgbd}), is then estimated as
\beq
\delta \Gamma (B_d)={G_F^2 m_b^3 f_B^2 |V_{cb}|^2 \over 8 \pi}\, |h_T-{5
\over 3} \, h_S|^2 \approx 0.009 \,\left ( {f_B \over 200 \, MeV }
\right )^2 \, ps^{-1}~,
\label{dgbdr}
\eeq
which constitutes about 1.5\% of the total decay rate.

\section{Discussion and Summary}
The estimates of the previous section, with all the uncertainty arising
from the present poor knowledge of the hadronic matrix elements,
demonstrate that an introduction of a new `centiweak' interaction,
suggested in eq.(\ref{lcw}), might slightly modify theoretical estimates
of the observable effects in $b$ hadron decays, possibly putting them
closer to the current experimental data. An additional reduction of the
ratio $\tau(\Lambda_b)/\tau(B_d^0)$ by about 5\% would place the
theoretical prediction within less than $2 \, \sigma$ from the
experimental number. An increase by about 15\% of the overall
nonleptonic decay rate of the $b$ hadrons might be helpful in
understanding the relatively low experimental value for the semileptonic
branching ratio $B_{sl}(B)$. Moreover this increase comes from a decay
channel with a single $c$ quark, thus effectively diluting the branching
ratio for the channel $b \to c \, {\overline c} \, s$ and thus somewhat
{\it reducing} the average charm yield $n_c$. This is contrary to the
analyses, where the enhancement of the total nonleptonic decay is
achieved by increasing the rate in the channel $b \to c \, {\overline c}
\, s$, thereby effectively correlating a low $B_{sl}(B)$ with a larger,
than experimentally observed, value of $n_c$. (For the most recent
review of this issue see Ref.\cite{nu}.)

As discussed, simultaneously with enhancing the decay of $\Lambda_b$ the
new interaction additionally suppresses the decay rate of $B^\pm$ with a
`sample' magnitude of the suppression being about 3\%. Combined with
about 1.5\% enhancement of the decay of $B_d$, this effect might
increase the ratio $\tau(B^\pm)/\tau(B_d)$ by about 4\%. This is to be
compared with the current experimental number \cite{pdg}:
$\tau(B^\pm)/\tau(B_d)=1.06 \pm 0.03$ and the `preferred' theoretical
values: $\tau(B^\pm)/\tau(B_d)-1 \approx 0.04$ \cite{mv} and
$\tau(B^\pm)/\tau(B_d)-1 \approx 0.05$ \cite{nu} (both theoretical
numbers are normalized to $f_B = 200 \, MeV$). Therefore, if with
improved experimental data this ratio of lifetimes moves toward the
upper edge of the currently allowed range, this can be interpreted as a
manifestation of the discussed new interaction.

The slightly more than one percent enhancement of the decay rate of the
$B_d$ meson, might seem insignificant. However, it should be noticed
that the standard theoretical analysis places this decay rate well
within 1\% from the average decay rate of the $B_s$ mesons:
$|\tau(B_d)/\tau(B_s)-1| <0.01$. Thus an experimental observation of a
breaking of this `one percent rule' may serve as another signal of the
new interaction.

Naturally, a speculation about a new four-quark interaction necessarily
invites concerns about its compatibility with other known phenomenology,
in particular with the limits on the `flavor changing neutral currents'
(FCNC). The interaction in eq.(\ref{lcw}) can in principle induce,
through interference with the standard weak interaction, effective FCNC
in two neutral channels: $b \, {\overline d}$ and $c \, {\overline u}$.
Thus there can appear a  contribution to the $B_d - {\overline B}_d$ and
$D^0 - {\overline D}^0$ oscillations. However, for the $B_d - {\overline
B}_d$ case, the transition $b\, {\overline d} \to c \, {\overline u}$
involves a right-chiral $u$ quark, which effectively decouples from the
weak interaction, up to effects proportional to the tiny $m_u$.
Therefore the contribution of the new interaction to the $B_d -
{\overline B}_d$ oscillations is practically nonexistent. The
`centiweak' transition $c \, {\overline u} \to b \, {\overline d}$
involves a right handed $b$ quark and thus can be followed by the weak
transition $b {\overline d} \to u \, {\overline c}$ (proportionally to
$m_b \, V_{bu} \, V_{cd}$). A parametric estimate of this contribution
to the $D^0 - {\overline D}^0$ oscillation rate is
\beq
\delta m_D \sim (G_F^2/\pi^2) \, |h_A| \, m_c^2 \, m_b \, |V_{bu}| \,
|V_{cd}| \, f_D^2 \sim 10^{-3} \, \Gamma(D^0)~,
\label{dmd}
\eeq
which is well below the current upper limit on $\delta m_D/\Gamma(D^0)$.

The compatibility of the `centiweak' interaction with the current limits
on FCNC, which is argued here for the specific flavor structure shown in
eq.(\ref{lcw}), may become invalidated if other flavor combinations are
added in the same manner. As is already mentioned, it might appear
`unnatural' to assume that the considered `centiweak' interaction is
significant only for the specific flavors as suggested by
eq.(\ref{lcw}). However on the technical side such assumption is
perfectly legitimate. The considered interaction is introduced here
purely phenomenologically, leaving aside possible reasons for its origin
within more general models of a `New Physics', since the main goal of
the present paper is to demonstrate the mechanism, allowing an
interference of a new interaction with the standard one due to the
breaking of chirality of the light quarks in hadrons. The only obvious
reference to the origin of the discussed interaction, made here, is that
it can arise from an exchange of an SU(2) singlet scalar $\phi$ with
color properties of a diquark. Clearly, an introduction of a massive
SU(2) singlet scalar field in the Standard Model does not affect the
precision tests of the electroweak theory. Given that the strength of
the induced four quark interaction is of order $10^{-2} \, G_F$, the
mass of the $\phi$ can easily be as high as in the several TeV range
without the need of assuming its large Yukawa couplings.

In summary. It is demostrated that an explanation of the apparently low
value of the ratio of the lifetimes $\tau(\Lambda_b)/\tau(B_d)$ as an
effect of new `centiweak' four-quark interaction, is at least a `logical
possibility', which does not appear to be in conflict with other known
phenomenology. Moreover, a moderate additional contribution from the new
interaction to the nonleptonic decay rate of a $b$ quark may be helpful
in understanding a somewhat low semileptonic branching ratio
$B_{sl}(B)$. For the $B^\pm$ and $B_d^0$ mesons the model predicts
additonal effects in the lifetimes, which are still within the
experimental errors, but which might become testable with more precise
data.

\section{Acknowledgements}
I acknowledge a warm hospitality of the Aspen Center for Physics, where
preliminary estimates for this paper were done. This work is supported
in part by DOE under the grant number DE-FG02-94ER40823.

\end{document}